\documentclass[conference]{IEEEtran}
\IEEEoverridecommandlockouts

\usepackage{amsmath,amsfonts,amssymb}
\usepackage{graphicx}
\usepackage{subcaption}
\usepackage{cite}
\usepackage{authblk, xcolor}

\title{Federated Learning for Semantic Communications over Multiple Access Channels\thanks{This research was supported in part by ONR grant N00014-24-1-2730 and Virginia Commonwealth Cyber Initiative (CCI).}}

\author[1]{Yalin E. Sagduyu}
\author[1]{Tugba Erpek}
\author[2]{Yi Shi}

\affil[1]{\normalsize  Nexcepta, Gaithersburg, MD, USA}

\affil[2]{\normalsize  Virginia Tech, Arlington, VA, USA}

\begin{document}

\bstctlcite{IEEEexample:BSTcontrol}

\maketitle
\begin{abstract}
This paper presents a federated learning (FL) framework for semantic communication (SemCom) over the multiple access channel (MAC), where multiple distributed transmitters simultaneously convey semantic information to a common receiver using end-to-end learned semantic representations. Semantic feature extraction, compression, wireless transmission, and semantic inference are jointly optimized through FL-based SemCom. Unlike conventional SemCom systems that train users independently, simultaneous semantic transmissions over the MAC enable concurrent training of multiple transmitters within the same transmission interval, thereby reducing training overhead and channel usage. Federated semantic aggregation is incorporated with shared and personalized encoder components, allowing transmitters with limited local training data to benefit from semantic representations learned from transmitters possessing larger datasets while preserving local adaptation. Evaluated using the CIFAR-10 dataset over fading channels, the proposed approach maintains high semantic inference accuracy under shared wireless access and significantly improves performance under severe data imbalance. In the meantime, simultaneous semantic transmissions can significantly reduce training overhead while preserving robust semantic performance across distributed transmitters with heterogeneous local datasets. These results highlight the potential of combining FL and SemCom to support scalable and communication-efficient edge intelligence in future wireless networks.
\end{abstract}

\section{Introduction}

Emerging wireless systems are expected to support machine learning-driven applications such as V2X, AR/VR, distributed sensing, and collaborative autonomy, requiring efficient transmission of semantic information rather than raw bits alone. Unlike conventional communication systems that separately perform source coding, channel coding, modulation, and receiver processing, semantic communication (SemCom) jointly learns semantic feature extraction, compression, wireless robustness, and task inference in an end-to-end manner \cite{guler2018semantic, xie2021deep, gunduz2022beyond, uysal2022semantic, sagduyu2024will, sagduyu2024joint}. 

SemCom further extends from point-to-point links to the multiple access channel (MAC) \cite{zhang2023deepma, li2023non, shen2025semantic, sagduyu2026MAC}. In a MAC, multiple distributed users simultaneously transmit information to a common receiver over shared wireless resources, making it a fundamental communication model for uplink data collection and collaborative edge intelligence applications. Such shared-access communication naturally arises in federated learning (FL), where multiple edge devices periodically exchange information with a common aggregation server to collaboratively train machine learning models without sharing raw data.

FL is essential to distributed wireless systems, where edge devices collaboratively optimize shared models without centralized access to local data \cite{chen2020joint, amiri2020federated, niknam2020federated}. Integrating FL with SemCom \cite{xu2024federated, nguyen2024efficient, peng2025personalized} offers a scalable framework for distributed semantic intelligence while preserving data locality, but introduces challenges associated with shared wireless resources, heterogeneous local datasets, and device coordination.

Existing FL-based SemCom studies primarily consider point-to-point communication or assume independent user transmissions, without explicitly accounting for simultaneous semantic transmissions over shared wireless channels. In contrast, practical uplink wireless systems naturally employ multiple access communication, where multiple users concurrently access the same channel. Supporting FL-based SemCom over the MAC therefore requires jointly learning semantic representations that remain robust to multi-user interference while efficiently leveraging heterogeneous local datasets through collaborative optimization. These challenges motivate the development of an integrated FL and SemCom framework tailored for shared wireless access.

This paper presents a SemCom framework for the MAC in which multiple distributed users simultaneously transmit semantic latent representations to a common receiver over a shared fading channel. Simultaneous semantic transmission reduces training epochs and channel uses, while FL with semantic aggregation enables transmitters with limited training samples to sustain strong semantic inference by leveraging representations learned from those with larger datasets. The proposed framework combines shared and personalized semantic encoder components to balance collaborative learning and local adaptation, allowing efficient semantic knowledge transfer while accommodating heterogeneous data distributions across transmitters. The results presented in this paper show that the proposed approach achieves robust semantic inference performance under shared wireless access and substantial gains in scenarios with severe training-data imbalance.

The remainder of the paper is organized as follows. Section~\ref{sec:semcommac} presents the SemCom framework over the MAC. Section~\ref{sec:FL} describes the FL training approach with shared and personalized
semantic encoders. Section~\ref{sec:perf} evaluates the performance. Finally, Section~\ref{sec:conclusion} concludes the paper.

\section{SemCom in Multiple Access Channel \label{sec:semcommac}}

\begin{figure}[t!]
    \centering
    \includegraphics[width=1.0\linewidth]{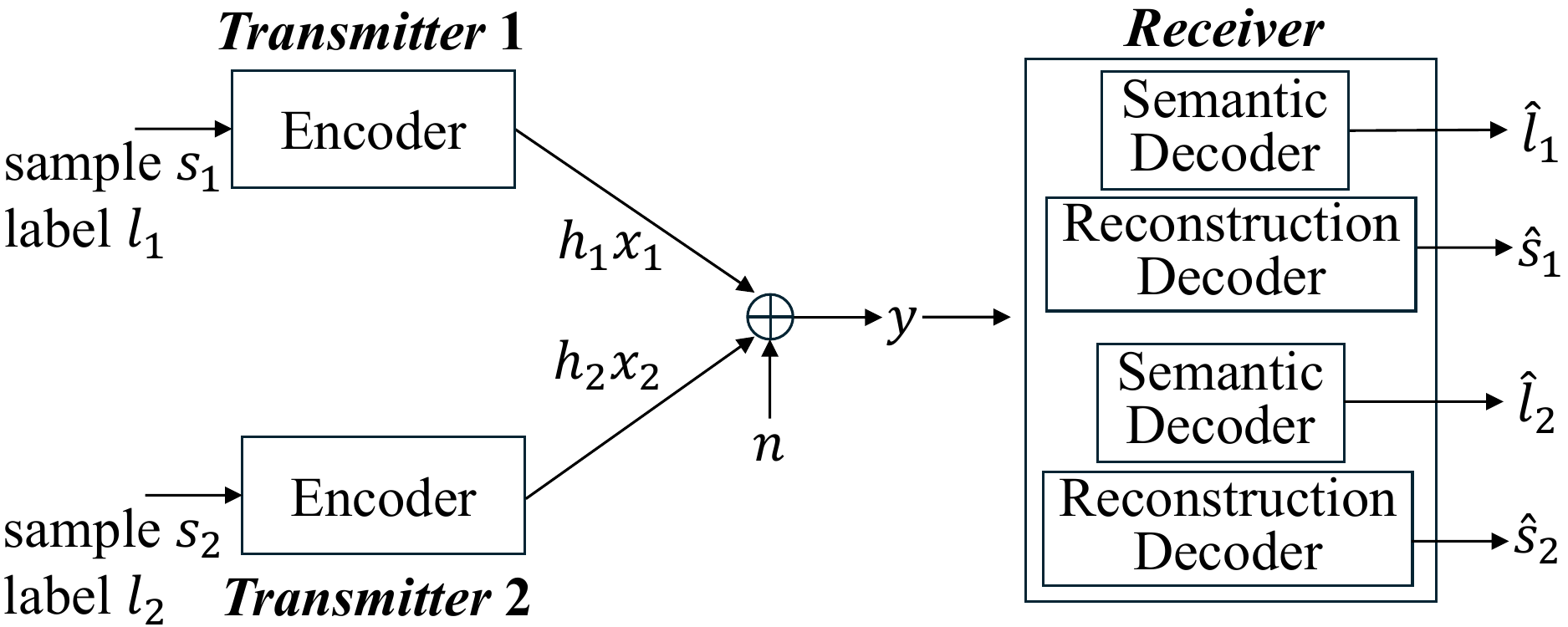}
    \caption{SemCom system model over the MAC.}
    \label{fig:system_model}
\end{figure}

The proposed framework considers distributed transmitters simultaneously transmitting semantic information to a common receiver over a shared fading wireless channel, as shown in Fig.~\ref{fig:system_model}. For each input image \(s_k\in\mathbb{R}^{C\times H\times W}\) with label \(l_k\), transmitter \(k\) applies a semantic encoder \(f_{\theta_k}(\cdot)\) to generate a latent semantic representation
\begin{equation}
x_k=f_{\theta_k}(s_k),
\end{equation}
where \(x_k\) corresponds to \(N\) complex wireless channel symbols (\(N\) denotes the latent dimension). For CIFAR-10, \(C=3\) and \(H=W=32\).

The semantic encoder consists of four convolutional layers organized into two convolutional blocks. Each convolutional layer uses \(3\times3\) kernels followed by batch normalization and ReLU activation. Each block is followed by \(2\times2\) max pooling and dropout with probability \(0.25\), reducing the input image resolution by a factor of four while increasing the feature dimension from 64 to 128 channels. The extracted feature maps are flattened and passed through a shared fully connected layer with 1024 neurons, followed by a transmitter-specific fully connected projection layer that generates \(N\) complex semantic channel symbols. In the FL implementation, the convolutional backbone and shared fully connected layer form the globally shared encoder component, while the final projection layer remains transmitter-specific. This design enables transmitters to collaboratively learn common semantic features while preserving local adaptation to transmitter-specific data and channel conditions. The transmitted latent representation is power normalized prior to wireless transmission.

Transmitters simultaneously transmit semantic latent representations over the MAC. The received signal is given by 
\begin{equation}
y=\sum_{k=1}^{K} h_k x_k+n
\end{equation} 
for $K$ transmitters, where \(h_k\) is the Rayleigh fading coefficient from transmitter \(k\) to the receiver, and \(n\) is the additive white Gaussian noise. The receiver first applies automatic gain control (AGC) to compensate for channel-dependent amplitude variations before semantic decoding. The normalized superposed semantic representation is then jointly processed by the semantic classification and reconstruction decoders, which share the same received latent representation while optimizing different semantic objectives. Unlike conventional receivers, no explicit multi-user detection, channel decoding, or signal separation is performed before semantic inference. For each transmitter \(k\), semantic classification and reconstruction decoders generate semantic predictions
\begin{equation}
\hat{l}_k=g_{\phi_k}^{(\mathrm{cls})}(y)
\end{equation}
and reconstructed images 
\begin{equation}
\hat{s}_k=g_{\psi_k}^{(\mathrm{rec})}(y),
\end{equation}
respectively. 
The semantic classification decoder is implemented as a fully connected neural network that takes the received \(2N\)-dimensional I/Q semantic representation as input and maps it through hidden layers with 512 and 256 neurons, each followed by ReLU activation and dropout with probability \(0.5\). A final fully connected layer outputs logits for the \(M\) semantic classes, where \(M=10\) for the CIFAR-10 dataset. The reconstruction decoder first expands the received semantic representation through fully connected layers of dimensions \(2N\!\rightarrow\!1024\!\rightarrow\!256(H/4)(W/4)\). The expanded representation is reshaped into a spatial feature tensor and processed by two transposed convolutional layers with 128 and 64 channels, each followed by batch normalization and ReLU activation. A final \(3\times3\) convolution followed by a sigmoid activation reconstructs the image with pixel values in \([0,1]\).

For transmitter \(k\), the semantic classification loss is defined as 
\begin{equation}
\mathcal{L}_k^{(\mathrm{cls})}=-\sum_{m=1}^{M} l_{k,m}\log \hat{l}_{k,m},
\end{equation}
where \(M\) is the number of semantic classes, \(l_{k,m}\) is the $m$-th element of the one-hot encoding of label \(l_k\), and \(\hat{l}_{k,m}\) is the predicted probability of class \(m\). For the original and reconstructed images\(s_k\) and \(\hat{s}_k\), the reconstruction loss is the mean squared error (MSE)  
\begin{equation}
\mathcal{L}_k^{(\mathrm{rec})}= \mathrm{MSE}(s_k,\hat{s}_k),
\end{equation}
\begin{equation}
\mathrm{MSE}(s_k,\hat{s}_k) = \frac{1}{CHW} \|s_k-\hat{s}_k\|_2^2.
\end{equation}
Peak Signal-to-Noise Ratio (PSNR) and Structural Similarity Index Measure (SSIM) are additionally included to improve image reconstruction quality. PSNR is defined as
\begin{equation}
\mathrm{PSNR}=10\log_{10}\!\left(\frac{\mathrm{MAX}^2}{\mathrm{MSE}}\right),
\end{equation}
where \(\mathrm{MAX}\) denotes the maximum possible pixel value (equal to 1 for the normalized images). SSIM is defined as
\begin{equation}
\mathrm{SSIM}(s_k,\hat{s}_k)=
\frac{(2\mu_s\mu_{\hat{s}}+C_1)(2\sigma_{s\hat{s}}+C_2)}
{(\mu_s^2+\mu_{\hat{s}}^2+C_1)(\sigma_s^2+\sigma_{\hat{s}}^2+C_2)},
\end{equation}
where \(\mu_s\) and \(\mu_{\hat{s}}\) are image means, \(\sigma_s^2\) and \(\sigma_{\hat{s}}^2\) are variances, \(\sigma_{s\hat{s}}\) is the covariance, and \(C_1 = 10^{-4}\) and \(C_2 = 9 \times 10^{-4}\) are the stability constants.

The loss for transmitter $k$ is given by  
\begin{equation}
\mathcal{L}_k=\sum_{j}\lambda_j\mathcal{L}_k^{(j)},
\end{equation}
where \(j\in\{\mathrm{cls},\mathrm{rec},\mathrm{ssim},\mathrm{psnr}\}\),  \(\mathcal{L}_k^{(\mathrm{ssim})}=1-\mathrm{SSIM}(s_k,\hat{s}_k)\) and \(\mathcal{L}_k^{(\mathrm{psnr})} \!\!= \!\!-\mathrm{PSNR}/10\), and \(\lambda_j\) is the corresponding weighting coefficient that controls the tradeoff between semantic inference accuracy and image reconstruction quality, allowing prioritization of downstream semantic task performance or reconstruction fidelity depending on application requirements.

The overall objective is 
\begin{equation}
\mathcal{L}_{\mathrm{total}}=\sum_{k=1}^{K} w_k\mathcal{L}_k,
\end{equation}
where \(w_k\) is the priority coefficient associated with transmitter \(k\). In the numerical results, we set \(w_k=1/K\). Simultaneous transmitter participation jointly optimizes multiple SemCom links, reducing communication epochs and channel uses compared to separately trained SemCom systems.

\section{FL for SemCom over Multiple Access Channel \label{sec:FL}}

SemCom over the MAC is trained with FL in which transmitters collaboratively optimize semantic encoders without centralized access to local data. The encoder parameters previously denoted by \(\theta_k\) are partitioned into globally shared and personalized components according to
\begin{equation}
f_{\theta_k}(s_k)=f_{\theta_k^{(p)}}^{(p)}(f_{\theta^{(g)}}^{(g)}(s_k)),
\end{equation}
where \(\theta^{(g)}\) denotes the globally shared encoder parameters corresponding to the convolutional backbone and shared fully connected layer, while \(\theta_k^{(p)}\) denotes the transmitter-specific projection layer parameters. This partition follows the encoder architecture described in Section~\ref{sec:semcommac}, enabling federated aggregation over common semantic feature extraction layers while keeping the final semantic projection personalized at each transmitter. The proposed architecture also reduces the communication and storage overhead of FL by restricting personalization to a single projection layer, while the computationally dominant convolutional feature extractor is collaboratively optimized across all transmitters through federated aggregation. In each FL round, transmitters independently update local semantic encoders according to 
\begin{equation}
\theta_k\leftarrow\theta_k-\eta_{\mathrm{enc}}\nabla_{\theta_k}\mathcal{L}_k,
\end{equation}
where \(\eta_{\mathrm{enc}}\) is the encoder-side learning rate.
The proposed partition reduces the communication overhead of FL since only the shared encoder parameters participate in federated aggregation, while the personalized projection layer remains local. Consequently, the framework simultaneously learns a common semantic feature space across distributed transmitters while allowing each transmitter to adapt its latent representation to its own data distribution, improving robustness under heterogeneous local datasets.

 After local optimization, transmitters transmit globally shared encoder parameters to the receiver, while personalized layers remain local. Global aggregation is performed using weighted federated averaging according to 
\begin{equation}
\theta^{(g)}\leftarrow\sum_{k=1}^{K}\frac{n_k}{\sum_{j=1}^{K}n_j}\theta_k^{(g)},
\end{equation}
where \(n_k\) is the number of training samples at transmitter \(k\). 

After local optimization, only the globally shared encoder parameters are transmitted to the receiver for federated aggregation, while personalized encoder layers remain local to each transmitter. The aggregated global encoder is then broadcast back to all transmitters for the next FL round. During each round, the receiver jointly processes the simultaneously received semantic latent representations from all transmitters over the MAC and updates the shared semantic classification and reconstruction decoders using the aggregated semantic objective. Let \(\phi=\{\phi_k\}_{k=1}^{K}\) and \(\psi=\{\psi_k\}_{k=1}^{K}\) denote the collections of classification and reconstruction decoder parameters for all receiver branches. Following aggregation, the receiver updates these decoder parameters according to
\begin{equation}
\phi,\psi\leftarrow\phi,\psi-\eta_{\mathrm{dec}}\nabla_{\phi,\psi}\mathcal{L}_{\mathrm{total}},
\end{equation}
where \(\eta_{\mathrm{dec}}\) is the receiver-side learning rate. Decoder updates are performed using simultaneous semantic latent transmissions over the shared fading MAC.

During training, transmitters simultaneously transmit semantic latent representations over the shared fading MAC while the receiver jointly processes the received superposed signal to generate semantic outputs for all transmitters. Personalized encoder layers remain transmitter-specific throughout FL training while still benefiting from globally shared semantic representations. The shared encoder backbone captures semantic features common across distributed transmitters, while personalized layers preserve transmitter-specific adaptation under heterogeneous local training data conditions.

\section{Performance Evaluation \label{sec:perf}}
This framework is evaluated using CIFAR-10 data of $32 \times 32 \times 3$ color images from 10 classes. Semantic encoders and decoders are optimized using the Adam optimizer with mini-batch stochastic gradient descent over 100 FL rounds. Batch normalization and dropout are employed in the networks to improve convergence and mitigate overfitting. Semantic classification accuracy, PSNR, and SSIM are evaluated under different signal-to-noise ratio (SNR) and $n_k$ values. The loss weights are set to \(\lambda_{\mathrm{cls}}=1\), \(\lambda_{\mathrm{rec}}=5\), \(\lambda_{\mathrm{SSIM}}=1\), and \(\lambda_{\mathrm{PSNR}}=1\). Learning rates \(\eta_{\mathrm{enc}}\) and  \(\eta_{\mathrm{dec}}\) are set to $10^{-3}$.

For $K = 2$ transmitters, Table~\ref{tab:homogeneous_results} shows that FL over the MAC achieves high accuracy for both transmitters comparable to standalone SemCom trained using the same large training data (namely, $n_1 = n_2 = 50$K), with only modest degradation in reconstruction quality. Simultaneous transmitter training within the same transmission interval reduces the number of transmitter-to-receiver channel uses by one half. These results indicate that multiple semantic streams can be jointly learned and decoded over a shared wireless channel without significantly compromising semantic inference performance. Although the receiver observes only a superposition of transmitted semantic representations, the learned semantic encoders and decoders effectively separate task-relevant information, preserving classification accuracy while maintaining acceptable reconstruction fidelity.

Table~\ref{tab:results} shows that under severe training data imbalance, e.g., with $n_1 = 1$K and $n_2 = 50$K local training samples, the transmitter with limited training dataset can still achieve strong semantic inference performance through federated semantic aggregation with the transmitter possessing larger dataset. FL over the MAC improves the accuracy of the transmitter with only 1K local training samples by approximately 87\%-104\% across the considered SNR and \(N\) settings compared to standalone SemCom training. The largest gains are observed for the data-constrained transmitter, demonstrating that federated semantic learning successfully transfers semantic knowledge acquired from richer datasets without requiring access to raw training samples. At the same time, the transmitter with large dataset experiences only a minor performance change, indicating that collaborative learning benefits weaker participants without significantly penalizing stronger ones.

Increasing the latent semantic dimension from \(N=10\) to \(N=100\) consistently improves semantic inference accuracy across all considered SNRs. The improvement is particularly evident for transmitters with limited local datasets, indicating that a higher-dimensional latent representation provides greater capacity for transferring semantic knowledge through federated aggregation while maintaining robust performance for transmitters with abundant training data.

\begin{table}[t]
\centering
\caption{SemCom performance (SNR $ = $ 20 dB, $N = $ 100).}
\label{tab:homogeneous_results}
\begin{tabular}{l|c|c|c|c}
\hline
Method & Accuracy & PSNR & SSIM & MSE \\
\hline
Single TX ($n_1 = 50$K) & 0.8214 & 20.54 & 0.7121 & 0.0103 \\
FL over MAC ($n_1 = 50$K) & 0.8239 & 19.40 & 0.6501 & 0.0134 \\
FL over MAC ($n_2 = 50$K) & 0.8255 & 19.40 & 0.6502 & 0.0135 \\
\hline
\end{tabular}
\end{table}

\begin{table}[t]
\centering
\caption{SemCom accuracy when training data sizes differ.}
\label{tab:results}
\begin{tabular}{l|l|c|c|c}
\hline
Method & Samples & SNR (dB) & Latent Dim. $N$ & Accuracy \\
\hline

Single TX & $n_1 =$1K & 20 & 10 & 0.3896 \\
Single TX & $n_2 =$50K & 20 & 10 & 0.8084 \\
FL over MAC  & $n_1 =$1K & 20 & 10 & 0.7353  \\
FL over MAC  & $n_2 =$50K & 20 & 10 & 0.7767\\
\hline

Single TX & $n_1 =$1K & 20 & 100 & 0.4001 \\
Single TX & $n_2 =$50K & 20 & 100 & 0.8214 \\
FL over MAC & $n_1 =$1K & 20 & 100 & 0.8034 \\
FL over MAC  & $n_2 =$50K & 20 & 100 & 0.8075 \\
\hline

Single TX & $n_1 =$1K & 10 & 10 & 0.3702 \\
Single TX & $n_2 =$50K & 10 & 10 & 0.7720 \\
FL over MAC & $n_1 =$1K & 10 & 10 & 0.6991 \\
FL over MAC  & $n_2 =$50K & 10 & 10 & 0.7192 \\
\hline

Single TX & $n_1 =$1K & 10 & 100 & 0.3851 \\
Single TX & $n_2 =$50K & 10 & 100 & 0.8135 \\
FL over MAC  & $n_1 =$1K & 10 & 100 & 0.7916 \\
FL over MAC & $n_2 =$50K & 10 & 100 & 0.8003 \\
\hline
\end{tabular}
\end{table}

\begin{figure}[t]
    \centering
    \includegraphics[width=0.92\linewidth]{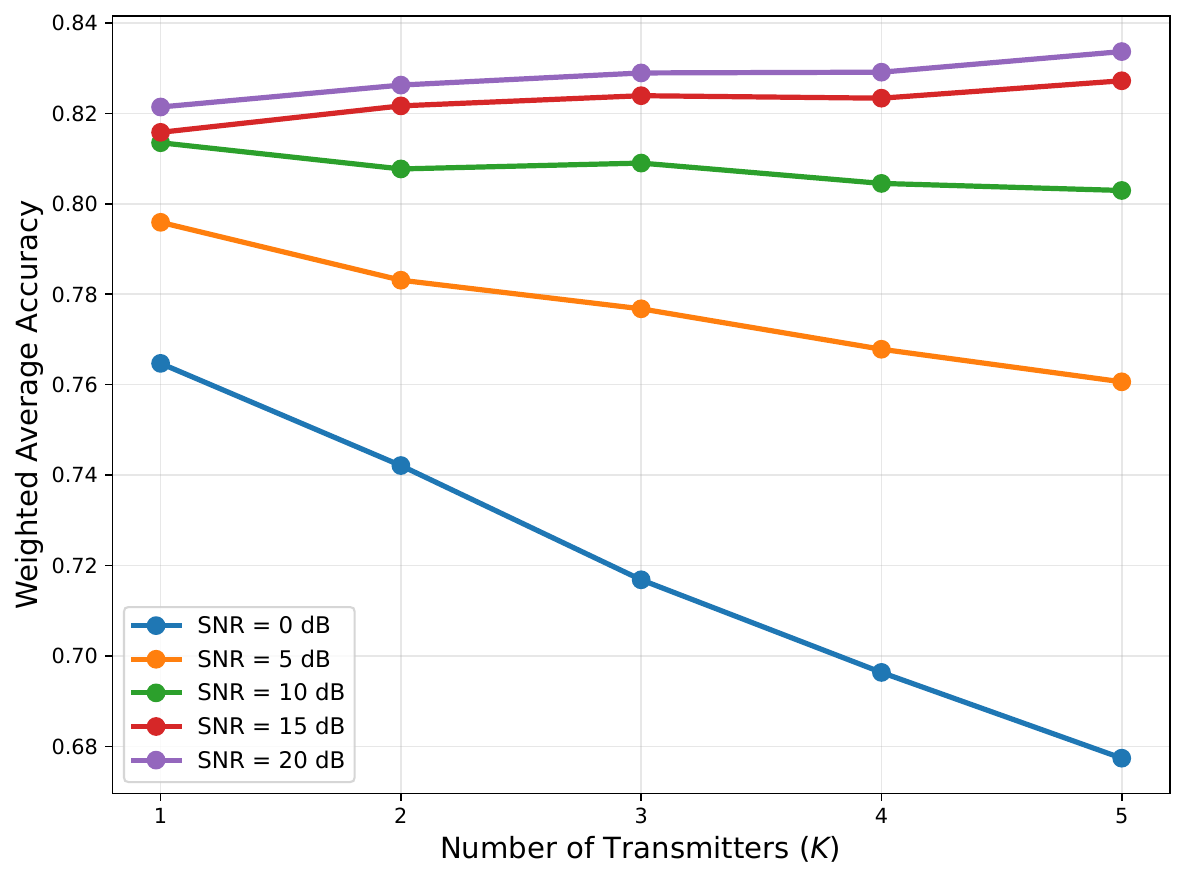}
    \caption{Average semantic classification accuracy versus the number of simultaneous transmitters under different SNR conditions.}
    \label{fig:acc_k}
\end{figure}

\begin{figure}[t]
    \centering
    \includegraphics[width=0.92\linewidth]{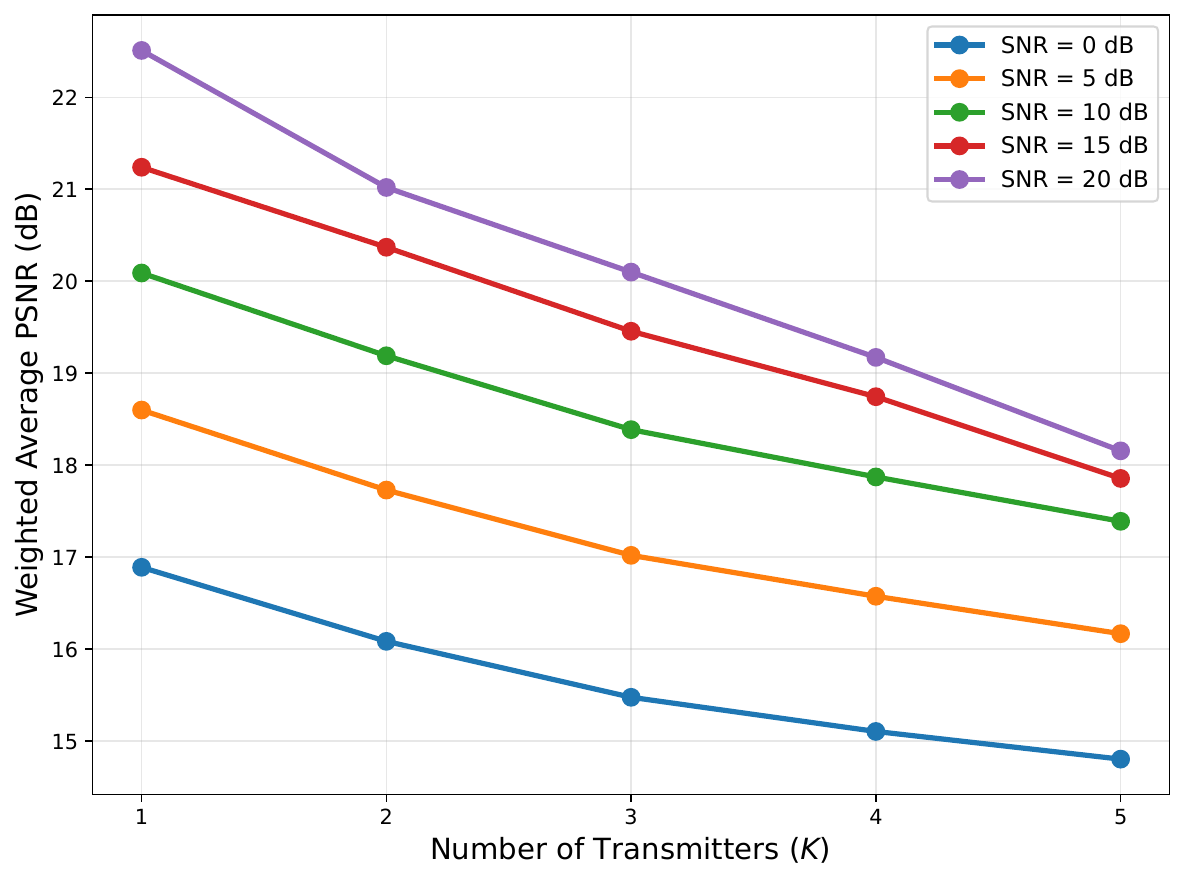}
    \caption{Average PSNR versus the number of simultaneous transmitters under different SNR conditions.}
    \label{fig:psnr_k}
\end{figure}

\begin{figure}[t]
    \centering
    \includegraphics[width=0.92\linewidth]{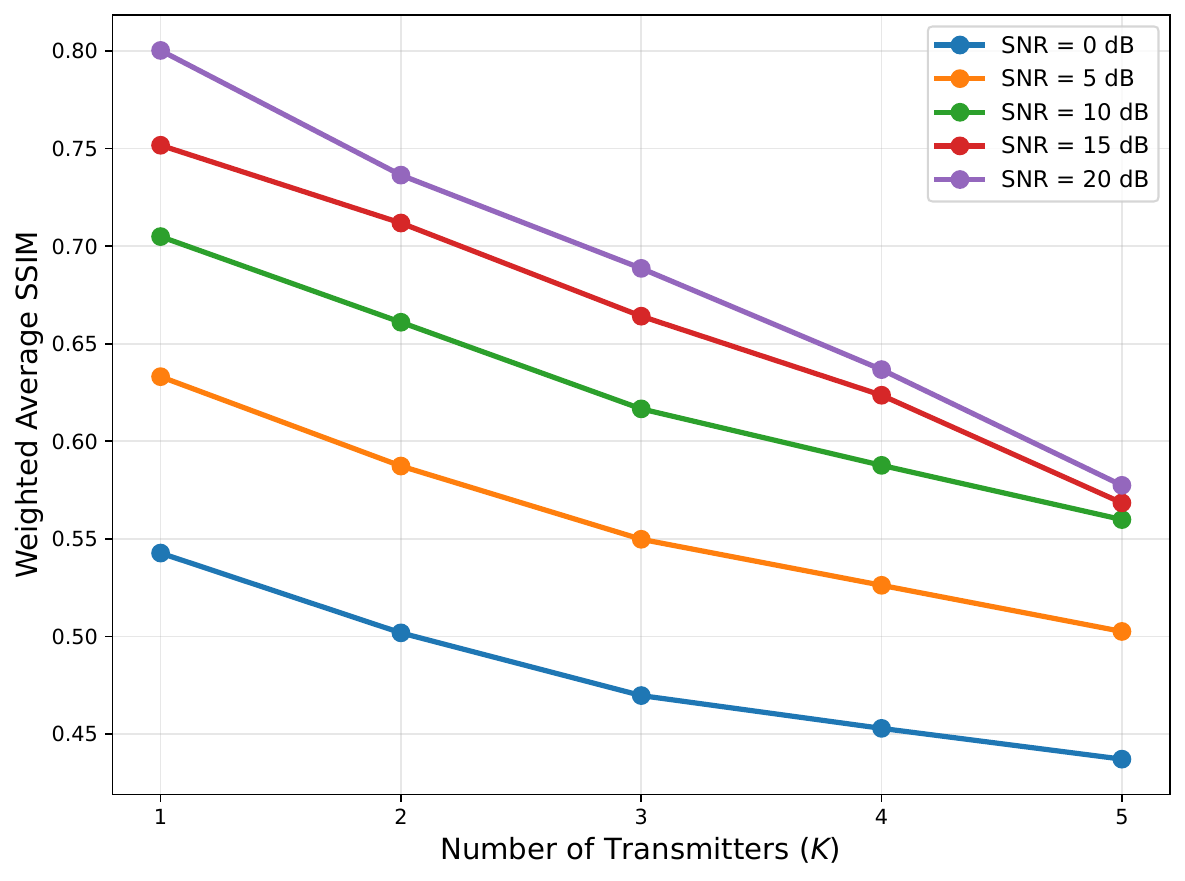}
    \caption{Average SSIM versus the number of simultaneous transmitters under different SNR conditions.}
    \label{fig:ssim_k}
\end{figure}

\begin{figure}[t]
    \centering
    \includegraphics[width=0.92\linewidth]{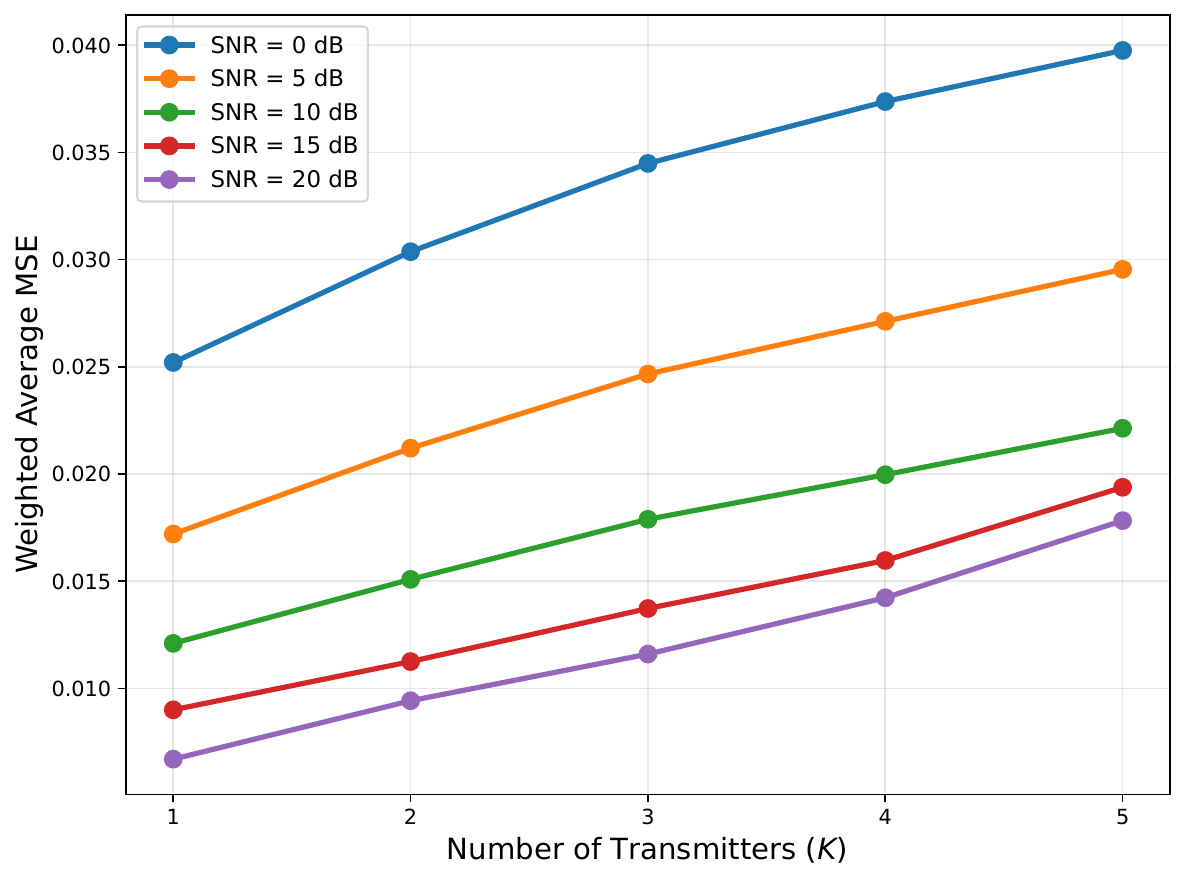}
    \caption{Average MSE versus the number of simultaneous transmitters under different SNR conditions.}
    \label{fig:mse_k}
\end{figure}

Figures~\ref{fig:acc_k}--\ref{fig:mse_k} evaluate the scalability of the proposed FL-based SemCom framework as the number of simultaneous transmitters, $K$, increases under different SNR conditions. In each case, the results are averaged uniformly among transmitters. Figure~\ref{fig:acc_k} shows the average semantic classification accuracy. As the number of simultaneous transmitters increases from \(K=1\) to \(K=5\), the impact on semantic classification accuracy depends on channel quality. The maximum change in average accuracy ranges from approximately \(1.5\%\) at \(20\) dB to \(11.4\%\) at \(0\) dB. Overall, these results demonstrate that the proposed SemCom framework scales well with the number of participating transmitters while maintaining robust semantic inference over shared MACs.

To emphasize image reconstruction, the reconstruction-related loss weights \(\lambda_{\mathrm{rec}}\), \(\lambda_{\mathrm{SSIM}}\), and \(\lambda_{\mathrm{PSNR}}\) can each be increased by a factor of five, while \(\lambda_{\mathrm{cls}}\) remains unchanged. Figure~\ref{fig:psnr_k} shows that the average PSNR gradually decreases with the number of simultaneous transmitters, indicating reduced pixel-level reconstruction fidelity. Likewise, Fig.~\ref{fig:ssim_k} shows a gradual reduction in average SSIM, reflecting lower structural similarity between the original and reconstructed images. Figure~\ref{fig:mse_k} exhibits the complementary trend, with the average MSE increasing as the number of transmitters grows due to stronger multiple access interference.

Across all figures, higher SNR consistently improves both semantic inference and reconstruction performance. As the number of simultaneous transmitters increases, image reconstruction quality degrades more noticeably than semantic classification accuracy. The relatively small variation in semantic classification accuracy suggests that semantic inference is more resilient to multiple access interference than image reconstruction. This behavior is consistent with the objective of SemCom, where preserving task-relevant semantic information is more important than faithfully reconstructing the source signal, enabling scalable support for large numbers of simultaneously transmitting users. Overall, the proposed FL-based SemCom framework scales well with the number of simultaneous transmitters while effectively preserving task-relevant semantic information over shared MAC.

\section{Conclusion \label{sec:conclusion}}

This paper presented an FL framework for SemCom over the MAC in which multiple distributed transmitters simultaneously convey semantic latent representations to a common receiver over shared fading channels. Simultaneous semantic transmissions over the MAC reduce channel uses and training overhead, while federated semantic aggregation substantially improves semantic inference for transmitters with limited local datasets under severe data imbalance. The proposed framework maintains high semantic inference accuracy under shared wireless access and effectively leverages knowledge learned across distributed transmitters to mitigate the impact of limited local training data. Semantic inference performance remains robust as the number of simultaneously transmitting users increases, highlighting the scalability of the proposed approach for multi-user wireless environments. By combining communication-efficient shared access with collaborative semantic learning, the proposed FL-enabled SemCom framework provides a promising solution for supporting distributed edge intelligence in future wireless systems that require efficient, scalable, and data-driven SemCom over shared wireless channels

\bibliographystyle{IEEEtran}
\bibliography{references}

\end{document}